%
\catcode`@=11 
\font\titlefontB=cmssdc10 at 18pt
\font\tmsm= zptmcm7t at 14pt
\font\tmsp= zptmcm7t at 12pt
\font\matdop= msbm10 
%
\font\bfsymbtext=cmmib10 
\font\bfsymbscript=cmmib10 at 7pt
\font\bfsymbscriptscript=cmmib10 at 5 pt
\def\boldsymbol#1{{\mathchoice%
	{\hbox{$\displaystyle\textfont1=\bfsymbtext\scriptfont1=%
		\bfsymbscript\scriptscriptfont1=\bfsymbscriptscript #1$}}%
	{\hbox{$\textstyle\textfont1=\bfsymbtext\scriptfont1=%
		\bfsymbscript\scriptscriptfont1=\bfsymbscriptscript #1$}}%
	{\hbox{$\scriptstyle\scriptfont1=\bfsymbscript%
		\scriptscriptfont1=\bfsymbscriptscript #1$}}%
	{\hbox{$\scriptscriptstyle%
	\scriptscriptfont1=\bfsymbscriptscript {#1}$}}}}
%
\catcode`@=12 
\def\ny#1{\boldsymbol#1}
%
%
%
%

%
%
%
\def\downnormalfill{$\,\,\vrule depth4pt width0.4pt
\leaders\vrule depth 0pt height0.4pt\hfill\vrule depth4pt width0.4pt\,\,$}
\def\WT#1{\mathop{\vbox{\ialign{##\crcr\noalign{\kern3pt}
      \downnormalfill\crcr\noalign{\kern0.8pt\nointerlineskip}
      $\hfil\displaystyle{#1}\hfil$\crcr}}}\limits}

%
%
%
%
%
%
\magnification = \magstep1
\hsize = 16.9 truecm
\vsize = 21.6 truecm 
\parindent=0.6 truecm 
\parskip = 2pt
\normalbaselineskip = 12pt plus 0.2pt minus 0.1pt
\baselineskip = \normalbaselineskip 
\topskip= 17pt plus2pt      
\voffset= 0 truecm   
\hoffset= -0.01 truecm      
%
%
%
%
\def\tfract#1/#2{{\textstyle{\raise0.8pt\hbox{$\scriptstyle#1$}\over%
\hbox{\lower0.8pt\hbox{$\scriptstyle#2$}}}}}
\def\mezzo{\tfract 1/2 }
\def\imezzi{\tfract i/2 }

\def\sesto{\tfract 1/6 }
\def\rad{\tfract 1/{\sqrt 2} }
\def\radik{\tfract 1/{\sqrt {k}} }
\def\radi2k{\tfract 1/{\sqrt {2k}} }
\def\der{\partial } 
\headline={\ifnum\pageno=0\line{ }\else {\vbox{\line{\hfil {\folio}}\line{\hrulefill}}}\fi}


\nopagenumbers
\count0=0

\null 

\line{\hfil IFUP-TH/2011-19}

\vskip 0.9 truecm 

\centerline {\titlefontB Field operators in topological quantum theories}

\vskip 2.5 truecm 

\centerline {\tmsm Enore Guadagnini}

\vskip 1.0 truecm 

\centerline{\sl Dipartimento di Fisica, Universit\`a di Pisa,}
\centerline{\sl and INFN, Sezione di Pisa,}
\centerline{\sl Largo B. Pontecorvo 3, 56127 Pisa, Italy}

\vskip 4.2 truecm 

\centerline {\bf Abstract}

\medskip

\midinsert \narrower 
 The perturbative approach to the topological quantum field theories of the Chern-Simons type formulated in $\hbox{\matdop R}^3$ is considered. By means of the canonical quantization  of the euclidean Chern-Simons lagrangian in the Landau gauge, a Fock space representation of the free field operators  is produced. The perturbative equivalence of the path-integral formalism and the field operator approach is exhibited.   
 The  expression of the one-loop effective action in background gauge is  derived. 
 \endinsert

\vfill\eject

\noindent {\tmsm 1 ~ Introduction}

\medskip

\noindent The introduction  of the  topological quantum field theories  of the Chern-Simons (CS) type presented in references [1, 2, 3] is based on  the path-integral formalism. The functional integral approach has also been used to guess the main topological properties of the   CS observables.  When the CS theory is formulated in $\hbox {\matdop R}^3$, explicit computations of the observables ---which can be identified with the expectation values of the Wilson line operators associated with  oriented links--- have been produced  by means of standard perturbation theory [4,5,6]. Now, in addition to the functional integral procedure,  perturbative quantum field theories also admit a canonical formulation  in which the field variables are represented by  operators acting on the linear space  of quantum states [7]. Precisely this field operator formulation of the perturbative  CS theory in $\hbox{\matdop R}^3$  is presented in the present  article. 
By means of the canonical quantization procedure applied to the gauge-fixed euclidean CS lagrangian in  $\hbox{\matdop R}^3$,   the  field operators are decomposed in terms of creation and annihilation operators acting on a  Fock space ${\cal F}$.  The field operator formulation of the CS theory is presented  in order to complete the conceptual setting concerning  this topological quantum fields model; the perturbative equivalence of the path-integral formalism and the field operator approach is exhibited.   Hopefully,   the basic aspects of the operator approach discussed here may find useful applications in the cases in which the standard perturbative approach cannot be used.  The fields  decomposition in terms of creation and annihilation operators is implemented in the  Landau gauge. On the other hand,  the use of the background gauge may turn to be useful for the computation of  certain topological observables. So,  the  expression of the one-loop CS effective action in background gauge is also derived. 
 
In order to make this article self-contained, a brief outlook on the field operator approach is presented in Section~2.   A  Fock space representation of the field operators  of the  euclidean CS theory in the Landau gauge is produced in Section~3.  
Finally, Section~4 contains the derivation of the expression of the one-loop  effective action of the CS theory in background gauge. 

\vskip 0.6 truecm 

\noindent {\tmsm 2 ~ Outlook on field operators}

\medskip

\noindent In order to describe the task of producing a  field operator formulation of a specific model of quantum field theory, let us firstly consider the general structure of the perturbative expansion. 

\bigskip 

\noindent {\tmsp 2.1 ~ Perturbative expansion}

\medskip

\noindent Let us consider the quantum field theories which admit a standard perturbative approach;  the perturbative expansion of the observables  is based on very few ingredients. Let the model of interest be characterized by 
the action $S[\phi ] = S_0[\phi ] + S_I[\phi]$, where  $\phi (x)$ indicates a set of real fields (for the sake of simplicity, let us assume that $\phi(x)$ represents  commuting variables). The so-called free action $S_0[\phi ] $  is a quadratic functional of the fields, $S_0[\phi ] = \mezzo   \int dx \,  \phi(x) \nabla \phi (x) $, in which $\nabla $ denotes a differential operator.  Whereas $ S_I[\phi]$ corresponds to the integral of the interaction lagrangian and contains cubic  and possibly quartic terms in powers of the fields. Let $\WT{\phi (x)\> \phi}(y) = 
i \, \nabla^{-1} (x,y) $ be the Feynman causal propagator where $\nabla^{-1} (x,y) $ stands for  a Green function of the $\nabla $ operator 
$$
\nabla \cdot  \nabla^{-1} (x,y) = \delta (x-y) \; . 
\eqno(1)
$$
The perturbative expansion of the expectation value $\langle F[\phi ] \rangle $ ---where the functional $F[\phi ]$  admits an expansion in powers of the fields---  is given by 
 $$
\langle F[\phi ] \rangle \equiv {\exp \left [  \imezzi \int dx dy {\delta \over \delta \phi (x)} \nabla^{-1} (x,y) {\delta \over \delta \phi (y)} \right ] \; F[\phi ] \; e^{i S_I [\phi ] } \Bigg |_{\phi =0}\over \exp \left [  \imezzi \int dx dy {\delta \over \delta \phi (x)} \nabla^{-1} (x,y) {\delta \over \delta \phi (y)} \right ] \;  e^{i S_I [\phi ] } \Bigg |_{\phi =0} }
\; . 
\eqno(2)
$$
Expression (2) indicates a formal sum of Feynman diagrams in which the so-called vacuum-to-vacuum diagrams  are eliminated. The sum of the vacuum-to-vacuum  diagrams, that is represented by the denominator of the ratio (2), factorizes in the numerator and cancels out with the denominator. So, there is no need to compute the vacuum-to-vacuum diagrams, which remain divergent even after the renormalization procedure has been introduced. The regularization/renormalization prescription and the definition of the composite field operators  play no role in the following discussion and  will be ignored. 

\bigskip 

\noindent {\tmsp 2.2 ~ Field theory formulations}

\medskip

\noindent  Expression (2) can be justified ---or derived--- by means of two different formulations of quantum field theories: the path-integral or the field operator methods. Both approaches eventually lead to expression (2) but, in the intermediate steps of the construction, within each method one needs to introduce a suitable set of assumptions. 

In the path-integral formulation, equation (2) takes the form 
$$
\langle F[\phi ] \rangle = {\int D \phi \; e^{i S [\phi ] } \, F[\phi ]  \over \int D \phi \; e^{i S [\phi ] } } \; .  
\eqno(3)
$$
A recent discussion on the use of the functional integration, for  perturbative and non-per\-tur\-ba\-tive  computations of the topological invariants of the CS theory, can be found in [8,9]. In the present article, the path-integral formulation will not be considered. 

The field operator approach consists of two parts: 

\item {(i)}  to each field component $\phi (x)$ one associates a   field operator   $  \widehat \phi (x)$  acting on a Hilbert space ${\cal F}$ that typically has a Fock space structure. $\widehat \phi (x)$ is a free field operator because it must satisfy the equation of motion $\nabla \widehat \phi (x) =0 $, which follows from the free action $S_0[\phi ]$. The equal-time algebra of the field operators is determined by the rules of the canonical quantization procedure; this simply means that the charges associated with the Noether currents  must be the generators of the corresponding symmetry transformations [10]. 

\item {(ii)} one finds a normalized vector $| 0 \rangle \in {\cal F}$, that corresponds to the vacuum state,   such that the  expectation value  of the time-ordered product of two field  operators on the vacuum state coincides with the Feynman propagator, 
$$
\langle 0 |\,  {\rm T} \left (  \widehat \phi (x) \widehat \phi (y)  \right ) | 0  \rangle = i \, \nabla^{-1} (x,y) \; .  
\eqno(4)
$$

\noindent When conditions (i) and (ii) are satisfied, equation (2) can be written as 
$$
\langle F[\phi ] \rangle = {\langle 0 |\,  {\rm T} \left ( \,  F[\, \widehat \phi \, ] \, e^{i S_I [\, \widehat \phi \, ] } \, \right ) | 0  \rangle \over \langle 0 |\,  {\rm T} \left ( \,  e^{i S_I [\, \widehat \phi \, ] } \, \right ) | 0  \rangle } \; .  
\eqno(5)
$$
In the following section it is shown how to provide points  (i) and (ii) with an explicit realization in the case of the euclidean topological quantum field theories of the CS type with a gauge-fixed lagrangian. 

\vskip 0.6 truecm 
 
\noindent {\tmsm 3 ~ Field operators in the CS model}

\medskip

\noindent Let us consider the CS quantum field theory [3]  in which the  gauge group is $SU(N)$, or any other compact simple Lie group $G$.  

\bigskip 

\noindent {\tmsp 3.1 ~ Essentials}

\medskip

\noindent When the model is defined in $\hbox{\matdop R}^3$, in the Landau gauge the action $S$  can be decomposed as $S= S_0 + S_I$ with 
$$
S_0 = \int d^3x \left  [ \, \mezzo \varepsilon^{\mu \nu \rho } A^a_\mu \partial_\nu A^a_\rho + \partial_\mu B^a A_\nu^a \, \eta^{\mu \nu } \sqrt \eta + \der_\mu {\overline c}^a \der_\nu c^a\,  \eta^{\mu \nu} \sqrt \eta \,  \right ] \; , 
\eqno(6)
$$
and 
$$
S_I = g \int d^3x \left  [  -  \sesto  \varepsilon^{\mu \nu \rho } \,  f^{abc }\,  A^a_\mu A_\nu^b A^c_\rho -  f^{abd} \, \der_\mu {\overline c}^a A^b_\nu c^d\,  \eta^{\mu \nu} \sqrt \eta \,  \right ] \; .  
\eqno(7)
$$
The real parameter $g$ is the coupling constant of the model; if the field variables are rescaled so that the coupling constant is placed in front of the whole lagrangian, in standard notation one has $k / 4 \pi = 1/ g^2$. 
The antisymmetric group tensor $f^{abc}$ denotes the structure constants of the Lie algebra of the gauge group $G$, and  the metric $\eta_{\mu \nu}$ in  $\hbox{\matdop R}^3$ is chosen to be ---in a cartesian coordinates system---  the $3\times 3$ identity  matrix. 
The gauge-fixed lagrangian  (6) and (7) defines an euclidean quantum field theory because  $\eta_{\mu \nu }$ is of euclidean type.  This choice for $\eta_{\mu \nu}$ guarantees that the topology  which is induced by the metric coincides with the standard topology of knot theory and with the standard 3-manifold topology.
The functional $S_0 + S_I$ is invariant under BRST transformations of the fields. The action of the BRST transformations  ---on the vector gauge fields $A_\mu^a(x)$, on the auxiliary field $B^a(x)$, on the ghost and antighost fields   $c^a(x)$ and ${\overline c}^a (x)$--- is given by  

$$\eqalign { 
& \delta_{G} \,  A^a_\mu (x) = \der_\mu c^a (x) - g f^{abd} A_\mu^b(x) c^d(x) \quad , \quad \delta_{G} \, c^a (x) = \mezzo g f^{abd} c^b (x) c^d (x)  \cr  & \delta_{G} \, {\overline c}^a(x) = - B^a (x) \quad , \quad  \delta_{G} \, B^a(x) = 0 \; . \cr} 
\eqno(8)
$$
The  conserved Noether current $J_\mu (x)$, which is associated with the BRST  transformations (8), assumes the form
$$
 J^\mu = \epsilon^{\mu \nu \rho} A_\nu^a \left ( D_\rho c\right )^a + B^a \left ( D^\mu c\right )^a  + \mezzo g f^{abd} \der^\mu {\overline c}^a c^b c^d \; ,  
 \eqno(9)
$$
in which $ \left ( D_\mu c\right )^a = \der_\mu c^a - g f^{abd}A_\mu^b c^d $. 
The energy-momentum tensor is 
$$\eqalign {
\Theta_{\mu \nu} &= \der_\mu B^a A_\nu^a + \der_\nu B^a A_\mu^a - \delta_{\mu \nu} \der^\rho B^a A_\rho^a \cr 
& \; \;  + \der_\mu {\overline c}^a \left ( D_\nu c\right )^a +  \der_\nu {\overline c}^a \left ( D_\mu c\right )^a - \delta_{\mu \nu}  \der^\rho {\overline c}^a \left ( D_\rho c\right )^a \; .  \cr }
\eqno(10)
$$
Only the gauge-fixing  lagrangian terms  depend on the metric $\eta_{\mu \nu}$, therefore  $\Theta_{\mu \nu}$ is equal to a BRST variation,  i.e. 
$$ 
\Theta_{\mu \nu} = \delta_G \left [ - \der_\mu {\overline c}^a(x) A_\nu^a - \der_\nu {\overline c}^a(x) A_\mu^a + \delta_{\mu \nu } \der^\rho  {\overline c}^a(x) A_\rho^a
 \right ] \; . 
 \eqno(11)
$$
Because of relation (11), the CS model is called a topological quantum field theory;  in facts property (11) implies that: 
\item {$\bullet$} the two-point correlator $\langle \Theta_{\mu \nu}  (x) \Theta_{\rho \sigma} (y) \rangle $ vanishes and, more generally,  the product of an arbitrary number of components of the energy momentum tensor has a vanishing expectation value,
$\langle \Theta_{\mu \nu}  (x_1) \Theta_{\rho \sigma} (x_2) \cdots   \Theta_{\lambda \tau} (x_n)\rangle =0 $. This means that, differently from ordinary field theories and differently from conformal models, the dynamical content of the CS theory is totally trivial;  
\item {$\bullet$} the expectation values of  metric-independent and BRST-invariant observables  are  topological invariants.  

\noindent The components of the Feynman propagator are given by [4]
$$\eqalign {  
\WT{A^a_\mu(x)\> A}{^b_\nu}(y) = - i \delta^{ab} \, \varepsilon_{\mu \nu \rho } \, {\partial \over \partial x_\rho } \Delta (x-y)  \quad &,  \quad 
   \WT{A^a_\mu(x)\> B}{^c}(y) =  i \delta^{ac} \, {\der \over \der x^\mu } \Delta (x-y) \cr 
 \WT{B^a(x)\> B}{^c}(y) = 0 \quad &, \quad  \WT{c^a(x)\> {\overline c}}{^b}(y) =  i \delta^{ab} \,  \Delta (x-y) \; . \cr }
\eqno(12)
$$
in which 
$$ 
\Delta (x) = \int {d^3 p \over (2 \pi )^3} {e^{ip_\mu x^\mu } \over p_\mu p^\mu} = {1\over 4 \pi | x | } \; . 
\eqno(13)
$$
By using the propagator components (12) and the interaction lagrangian (7), one can show that (to all orders of perturbation theory) the CS model is renormalizable [11].  The explicit  computation of the expectation values of the Wilson line operators has been produced at the third nontrivial order of perturbation theory [5,6]; the results are in complete agreement with the general properties [12] of the expectation values that follow from the symmetries of the model. Expressions (12) and (13) can also be understood as the basic ingredients for the construction of integral formulas for link invariants  [13]. 
 
 \bigskip 

\noindent {\tmsp 3.2 ~ Commuting variables}

\medskip

\noindent In order to follow the recipe of the canonical quantization procedure, one has to choose a time direction.  Let each element of $ \hbox{\matdop R}^3$ be identified by means of the real cartesian coordinates $x^\mu = (x^1, x^2, x^3)$, the third component $x^3$ will be interpreted as the ``time" component, $x^3 = t$. The components $x^i = (x^1, x^2) = x_i = (x_1, x_2) $ will be used to label the points of each spatial plane at fixed time.    Since 
$$
{\delta S_0 \over \delta (\partial_t A^a_1(x))} = A^a_2(x) \quad , \quad 
{\delta S_0 \over \delta (\partial_t B^a(x))} = A^a_3(x) \; , 
\eqno(14)
 $$
the field operators $\widehat A^a_\mu (x) $ and $\widehat B^a (x)$, which are associated with the commuting variables, must satisfy the canonical commutation relations   
$$
[\widehat  A^a_1(\boldsymbol {x} , t) , \widehat A^b_2(\boldsymbol {y} , t)] = i \delta^{ab} \, \delta (\boldsymbol {x} - \boldsymbol {y}  ) = [ \widehat B^a (\boldsymbol {x} , t) , \widehat A^b_3(\boldsymbol {y} , t)] \; ;   
\eqno(15)
$$
all the remaining equal-time commutators vanish. 
Given a couple $\widehat q$ and $\widehat p$ of canonically conjugated variables, $[ \, \widehat q \, , \,  \widehat p \, ] = i $, one can define the annihilation and creation operators $a = (\widehat q +i \, \widehat p \, ) / \sqrt 2 $ and $a^\dagger = (\widehat q - i \, \widehat p\, )/ \sqrt 2 $  that can be used to define a Fock space representation of the operators $\widehat q$ and $\widehat p$.  Similarly, the commutation relations (15) imply that a Fock space representation of the field operators can be produced. 

Let us assume that, on each spatial plane, the field  modes with definite values of the spatial  momentum are well defined,  
$$\eqalign { 
\widehat A^a_1 (x) &= \! \! \int {d^2k \over 2 \pi} \rad \bigg \{ \cos \theta_{\ny {k}} \left [ u_-^a(\boldsymbol {k}, t) e^{i\boldsymbol {k x}} + u_+^{a} ( \boldsymbol {k}, t) e^{-i \boldsymbol {k  x}} \right ]  \cr 
 & \qquad \qquad ~ - \sin \theta_{\ny {k}} \left [ -i u_-^a(\boldsymbol {k}, t) e^{i\boldsymbol {k  x}} + i u_+^{a  } ( \boldsymbol {k}, t) e^{-i \boldsymbol {k  x}} \right ] \bigg \} \; , \cr }
 \eqno(16)
 $$
\vskip -0.6 truecm 
$$\eqalign {
\widehat A^a_2 (x) &= \! \! \int {d^2k \over 2 \pi} \rad \bigg \{ \cos \theta_{\ny {k}} \left [ - i u_-^a(\ny {k}, t) e^{i\boldsymbol {kx}} + i u_+^{a  } ( \ny {k}, t) e^{-i \boldsymbol  {kx}} \right ]  \cr 
 & \qquad  \qquad ~~~  + \sin \theta_{\ny {k}} \left [ u_-^a(\ny {k}, t) e^{i\boldsymbol  {k  x}} +  u_+^{a}( \ny {k}, t) e^{-i \boldsymbol {kx}} \right ] \biggr \} \; , \cr }
 \eqno(17)
$$
\vskip -0.3 truecm
$$
\widehat A^a_3 (x) = \int {d^2k \over 2 \pi} \rad \bigg \{  - i v_-^a(\ny {k}, t) e^{i\boldsymbol  {k  x}} + i v_+^{a } ( \ny {k}, t) e^{-i \boldsymbol  {k  x}}  \bigg \} \; , 
 \eqno(18)
$$
$$ 
\widehat B^a (x) = \int {d^2k \over 2 \pi} \rad \bigg \{   v_-^a(\ny {k}, t) e^{i \boldsymbol  {k  x}} + v_+^{a}( \ny {k}, t) e^{-i \boldsymbol  {k  x}}  \bigg \} \; , 
 \eqno(19)
$$
where  $e^{i\boldsymbol  {k  x}} = \exp [ i (k_1 x_1 + k_2 x_2)] $ and 
$$ 
\cos \theta_{\ny {k}} = k_1 / k  \quad ,  \quad \sin \theta_{\ny {k}} = k_2 / k  \quad ,  \quad  k = \sqrt {k_1^2 + k_2^2} \; . 
\eqno(20)
$$
The $\cos \theta_{\ny {k}} $ and $\sin \theta_{\ny {k}}$ factors in expressions (16) and (17) have been introduced in order to make  the polarization choice for $\widehat A^a_1 (x)$ and $\widehat A^a_2 (x)$ agree with the directions ---on the spatial plane--- that are naturally defined by the values of the spatial momentum;    this just  simplifies the structure of the equations of motion. The validity of the following equal-time commutators 
$$\eqalign { 
&\left [ u_-^a(\ny {k}, t) , u_+^{b }( \ny {p}, t) \right ] = \delta^{ab}\, \delta (\ny {k} -  \ny {p}  ) = \left [ v_-^a(\ny {k}, t) , v_+^{b } ( \ny {p}, t) \right ]  \; , \cr  
&\left [ u_-^a(\ny {k}, t) , v_-^b ( \ny {p}, t)\right  ] =
\left [ u_-^a(\ny {k}, t) , v_+^{b  }( \ny {p}, t) \right ] =  0 \; , \cr 
&\left [ u_+^{a  } (\ny {k}, t) , v_-^b ( \ny {p}, t) \right ] = \left [ u_+^{a } (\ny {k}, t) , v_+^{b} ( \ny {p}, t) \right ] =0  \; ,  \cr } 
 \eqno(21)
$$
guarantees that  the canonical commutation relations (15) are satisfied. 

The time evolution of the operators is determined by the equations of motion that  must be derived from the free action (6),  
$$
\varepsilon^{\mu \nu \rho} \der_\nu \widehat A^a_\rho (x)  + \der^\mu \widehat B^a (x) = 0 \quad , \quad \der^\mu \widehat A^a_\mu (x) = 0  \; . 
\eqno(22)
$$
Equations (22) imply $\der_\mu \der^\mu \widehat A^a_\nu (x) = 0 $ and $\der_\mu \der^\mu \widehat B^a (x) = 0 $. 
By inserting the expressions (16)-(19) in equations (22), one finds  
$$\eqalign { 
{d\over dt } \, u_-^a (\ny {k} , t ) = - k  \, v_+^{a  } (- \ny {k} , t ) \quad &, \quad  {d\over dt} \, v_-^a (\ny {k} , t ) = - k  \, u_+^{a } (- \ny {k} , t ) \cr 
{d\over dt } \, u_+^{a } (\ny {k} , t ) = - k  \, v_-^{a} (- \ny {k} , t ) \quad &, \quad  {d\over dt} \, v_+^{a } (\ny {k} , t ) = - k  \, u_-^{a} (- \ny {k} , t ) \; . \cr }    
\eqno(23)
$$
The solution of the differential equations (23) is 
$$\eqalign {  
 u_-^a (\ny {k} , t ) &= u_-^a (\ny {k} ) \, {\rm Ch} (k t)  - v_+^{a  } (- \ny {k}  ) \, {\rm Sh} ( k t ) \; , \cr 
 v_-^a (\ny {k} , t ) &= v_-^a (\ny {k}  ) \, {\rm Ch} (k t)  - u_+^{a  } (- \ny {k}  ) \, {\rm Sh} ( k t ) \; , \cr 
  u_+^{a} (\ny {k} , t ) &= u_+^{a} (\ny {k}  ) \, {\rm Ch} (k t)  - v_-^a (- \ny {k}  ) \, {\rm Sh} ( k t ) \; , \cr 
 v_+^{a} (\ny {k} , t ) &= v_+^{a}(\ny {k}  ) \, {\rm Ch} (k t)  - u_-^a (- \ny {k}  ) \, {\rm Sh} ( k t )  \; , \cr } 
 \eqno(24)
 $$
 where $u_{\pm}^a (\ny {k}  )$ and  $v_{\pm}^a (\ny {k}  )$ denote the  mode operators  at time $t=0$ satisfying 
 $$\eqalign { 
&\left [ u_-^a(\ny {k } ) , u_+^{b}( \ny {p}  ) \right ] = \delta^{ab} \delta (\ny {k} -  \ny {p}  ) = \left  [ v_-^a(\ny {k} ) , v_+^{b} ( \ny {p } ) \right ]  \; , \cr 
&\left [ u_-^a(\ny {k } ) , v_-^b ( \ny {p } ) \right ] =
\left [ u_-^a(\ny {k } ) , v_+^{b}( \ny {p } )\right ] =  0 \; , \cr 
&\left [ u_+^{a} (\ny {k } ) , v_-^b ( \ny {p } ) \right  ] = \left [ u_+^{a} (\ny {k } ) , v_+^{b} ( \ny {p } ) \right  ] =0  \; .  \cr } 
 \eqno(25)
$$
Expressions  (24) and (25) provide equations (21) with  an explicit solution.  

 \bigskip 

\noindent {\tmsp 3.3 ~ Anticommuting variables}

\medskip

\noindent Let us now consider the operators $\widehat {c^a}(x)$ and $\widehat {{\overline c}^a}(x)$ which are associated with the anticommuting variables.  The  fields $ {c^a}(x)$ and ${{\overline c}^a}(x)$ correspond to independent variables and since 
$$
{\delta S_0 \over \delta (\partial_t c^a(x))} = \partial_t \, {\overline c}^a(x) \quad , \quad 
{\delta S_0 \over \delta (\partial_t {\overline c}^a(x))} = - \partial_t \,  c^a(x)
 \; , 
\eqno(26)
 $$
the canonical (anti-) commutation relations take the form 
$$
\left \{ \widehat {c^b} (\ny {x}, t) , \der_t\, \widehat {{\overline c}^a}(\ny {y} , t) \right \} = i \delta^{ab}\, \delta (\ny {x} - \ny {y}  )\; ,  
\eqno(27)
$$
and 
$$
\left \{  \widehat {{\overline c}^a}(\ny {y} , t) , 
\der_t\, \widehat {c^b} (\ny {x}, t) 
\right \} = - i \delta^{ab}\, \delta (\ny {x} - \ny {y } )\; . 
\eqno(28)
$$
The free equations of motion ---derived from expression (6)--- are 
$$ 
\der^\mu \der_\mu \widehat {c^a}(x) = 0 =  \der^\mu \der_\mu \widehat {{\overline c}^a}(x) \; . 
\eqno(29)
$$
Similarly to the case of the commuting variables,  the ghost and antighost field operators can be decomposed in modes with definite values of the spatial momentum. A simple representation  is given by 
$$ 
\widehat {c^a} (x) = \int {d^2k \over 2 \pi} \radik    \,  w^a(\ny {k} , t )  \, e^{i\boldsymbol  {k x}}   \; , 
 \eqno(30)
$$
$$ 
\widehat  {{\overline c}^a}(x)  = \int {d^2k \over 2 \pi} \radik \,   ( - i ) \,  z^a(\ny {k} , t )  \,  e^{i\boldsymbol  {k  x}}   \; .  
 \eqno(31)
$$
Equations (27) and (28) require 
$$
\left \{ w^a (\ny {k} , t)  , \der_t \,  z^b (\ny {p} , t) \right \} =  - k \, \delta^{ab}\, \delta (\ny {k} + \ny {p}  )  \quad , \quad \left \{ z^a ( \ny {k} , t)   , \der_t \,  
 w^b (\ny {p} , t)  \right \} =  k \, \delta^{ab}\, \delta (\ny {k} + \ny {p } )   \; .   
\eqno(32)
$$
A complete  solution of equations (29) and (32) is given by 
$$\eqalign { 
w^a (\ny {k} , t) &= w_-^a (\ny {k } )  \, {\rm Ch} (k t) + z_+^{a} (- \ny {k } )  \, {\rm Sh} ( k t ) \cr 
z^a (\ny {k} , t) &= z_-^a (\ny {k } ) \, {\rm Ch} (k t) - w_+^{a}(- \ny {k } ) \, {\rm Sh} ( k t ) \; , \cr } 
\eqno(33)
$$
with 
$$\eqalign { 
&\left \{ w_-^a(\ny {k } ) , w_+^{b}( \ny {p } ) \right \} = \delta^{ab} \, \delta (\ny {k} -  \ny {p } ) = \left \{ z_-^a(\ny {k } ) , z_+^{b} ( \ny {p } ) \right \}  \; , \cr  
&\left \{ w_-^a(\ny {k } ) , z_-^b ( \ny {p } ) \right \} =
\left \{ w_-^a(\ny {k } ) , z_+^{b}( \ny {p } ) \right \} =  0 \; , \cr 
&\left \{ w_+^{a} (\ny {k } ) , z_-^b ( \ny {p } ) \right \} = \left \{ w_+^{a} (\ny {k } ) , z_+^{b} ( \ny {p } ) \right \} =0  \; .  \cr }  
 \eqno(34)
$$

\bigskip 

\noindent {\tmsp 3.4 ~ Fock space representation of the field operators}

\medskip

\noindent The time evolution of the mode operators, which is  explicitly displayed in equations (24) and (33),  does not assume a diagonal form. If the operators  $ \{ u_{\pm}^a(\ny {k } ), v_{\pm}^{a} (\ny {k } ) \}$ and $\{ w_{\pm}^a (\ny {k } ), z_{\pm}^{a}(\ny {k } )\}$  are  interpreted as standard annihilation and creation operators that act  on a Fock space, the corresponding    ``vacuum" state $|  \omega \rangle $,  which is defined by the equations
$$\eqalign {  
u_-^a (\ny {k } )| \omega \rangle =0 \quad &, \quad v_-^a (\ny {k } )| \omega \rangle =0 \; , 
\cr 
w_-^a (\ny {k } )|  \omega \rangle =0 \quad &, \quad z_-^a (\ny {k } )| \omega \rangle =0 \; , \cr }
\eqno(35)
$$
\vskip - 0.3 truecm 
\noindent is not a stationary state; this implies that the mean values of the field operators computed with respect to $| \omega \rangle $ are not invariant under time translations. In fact, it turns out that   the expectation values of the time-ordered product of the field operators (16)-(19) and (30)-(31) on the state $| \omega \rangle $ differ from the expressions (12) of the Feynman propagators. 

In order to determine the annihilation and creation operators that canonically define the  Fock states space, let us consider the following combinations of the modes that diagonalize the free time evolution,  
$$\eqalign { 
& {\hskip -0.7 truecm} h_-^a (\ny {k } ) =  \rad \left [ u_-^a(\ny {k } ) + v_+^{a} (- \ny {k } )  \right ] \quad , \quad g_-^a (\ny {k } ) = \rad \left [ v_-^a(\ny {k } ) + u_+^{a} (- \ny {k } )  \right ] \cr 
& {\hskip -0.7 truecm} h_+^a (\ny {k } ) =  \rad \left [ u_+^{a} (\ny {k } ) - v_-^{a} (- \ny {k } )  \right ] \quad , \quad g_+^a (\ny {k } ) = \rad \left [ v_+^{a}(\ny {k } ) - u_-^{a} (- \ny {k } )  \right ] \cr 
& {\hskip -0.7 truecm} \lambda_-^a(\ny {k } ) = \rad \left [ w_-^a (\ny {k } ) - z_+^{a} (- \ny {k } ) \right ] \quad ,  \quad \xi_-^a (\ny {k } ) = \rad \left [ z_-^a(\ny {k } ) + w_+^{a} (- \ny {k } ) \right ] \cr 
& {\hskip -0.7 truecm} \lambda_+^a(\ny {k } ) = \rad \left [ w_+^{a} (\ny {k } ) - z_-^{a} (- \ny {k } ) \right ] \quad ,  \quad \xi_+^a (\ny {k } ) = \rad \left [ z_+^{a}(\ny {k } ) + w_-^{a} (- \ny {k } ) \right ] \; .     \cr } 
\eqno(36)
$$
It is convenient to define the operators  
$$\eqalign { 
& {\hskip -0.7 truecm} \alpha_-^a (\ny {k } ) =  \rad \left [ h_-^a(\ny {k } ) + g_-^{a} ( \ny {k } )  \right ] \quad , \quad \beta_-^a (\ny {k } ) = \rad \left [ h_-^a(\ny {k } ) - g_-^{a} ( \ny {k } )  \right ] \cr 
& {\hskip -0.7 truecm} \alpha_+^a (\ny {k } ) =  \rad \left [ h_+^{a  } (\ny {k } ) + g_+^{a} ( \ny {k } )  \right ] \quad , \quad \beta_+^a (\ny {k } ) = \rad \left [ h_+^{a  }(\ny {k } ) - g_+^{a} ( \ny {k } )  \right ] \; . \cr }      
\eqno (37)
$$
The following commutation or anticommutation  relations are satisfied 
$$\eqalign { 
&\left [ \alpha_-^a(\ny {k } ) , \alpha_+^{b}( \ny {p } ) \right ] = \delta^{ab} \delta (\ny {k} -  \ny {p } ) = \left  [ \beta_-^a(\ny {k } ) , \beta_+^{b} ( \ny {p } ) \right ]  \; , \cr 
&\left [ \alpha_-^a(\ny {k } ) , \beta_-^b ( \ny {p } ) \right ] =
\left [ \alpha_-^a(\ny {k } ) , \beta_+^{b}( \ny {p } )\right ] =  0 \; , \cr  
&\left [ \alpha_+^{a} (\ny {k } ) , \beta_-^b ( \ny {p } ) \right  ] = \left [ \alpha_+^{a} (\ny {k } ) , \beta_+^{b} ( \ny {p } ) \right  ] =0  \; ; \cr }    
 \eqno(38)
$$
$$\eqalign { 
&\left \{ \lambda_-^a(\ny {k } ) , \lambda_+^{b}( \ny {p } ) \right \} = \delta^{ab} \, \delta (\ny {k} -  \ny {p } ) = \left \{ \xi_-^a(\ny {k } ) , \xi_+^{b} ( \ny {p } ) \right \}  \; , 
\cr 
&\left \{ \lambda_-^a(\ny {k } ) , \xi_-^b ( \ny {p } ) \right \} =
\left \{ \lambda_-^a(\ny {k } ) , \xi_+^{b}( \ny {p } ) \right \} =  0 \; , \cr 
&\left \{ \lambda_+^{a} (\ny {k } ) , \xi_-^b ( \ny {p } ) \right \} = \left \{ \lambda_+^{a} (\ny {k } ) , \xi_+^{b} ( \ny {p } ) \right \} =0  \; .     \cr } 
 \eqno(39)
$$
The commutators (38) and (39)  have a canonical structure; so, let us consider a  standard representation of the creation and annihilation operators $\alpha^a_{\pm}, \beta^a_{\pm}$ and $\lambda^a_{\pm} , \xi^a_{\pm}$ with 
$$
  \left [ \alpha^a_- (\ny {k}) \right ]^\dagger =\alpha^a_+( \ny{k})  \quad , \quad \left [ \beta^a_- (\ny {k}) \right ]^\dagger  =  \beta^a_+( \ny{k})    \; ,       
\eqno(40)
$$
$$
  \left [ \lambda^a_- (\ny {k}) \right ]^\dagger =\lambda^a_+( \ny{k})  \quad , \quad \left [ \xi^a_- (\ny {k}) \right ]^\dagger  =  \xi^a_+( \ny{k})    \; .        
\eqno(41)
$$
Relations (36) and  (37) can be used to express the operators $u_{\pm}^a, v_{\pm}^a  $ and  $w_{\pm}^{a } , z_{\pm}^a$  in terms of $\alpha^a_{\pm}, \beta^a_{\pm}$ and $\lambda^a_{\pm} , \xi^a_{\pm}$; then, the field operators  (16)-(19)  take  the form 
$$\eqalign { 
\widehat A^a_1 (x) &= \int {d^2k \over 2 \pi} \rad \bigg \{ \left [  i \sin \theta _{\ny {k}}  \, \alpha_-^{a } ( \boldsymbol {k})  + \cos \theta_{\ny {k}}  \, \beta_-^a(\boldsymbol {k})   \right ]  e^{i\boldsymbol {k x} - k t }     \cr 
 & \qquad \qquad ~ + \left [ \cos \theta_{\ny {k}}  \, \alpha_+^a(\boldsymbol {k})  - i \sin \theta _{\ny {k}} \,  \beta_+^{a } ( \boldsymbol {k} )  \right ]  e^{- i\boldsymbol {k x} + k t }   \bigg \} \; , \cr } 
 \eqno(42)
$$
\vskip -0.6 truecm
$$\eqalign { 
\widehat A^a_2 (x) &= \int {d^2k \over 2 \pi} \rad \bigg \{ \left [ -i \cos \theta_{\ny {k}}  \, \alpha_-^a(\boldsymbol {k})  +  \sin \theta _{\ny {k}}  \, \beta_-^{a } ( \boldsymbol {k})  \right ]  e^{i\boldsymbol {k x} - k t }      \cr 
 & \qquad \qquad ~ + \left [ \sin \theta_{\ny {k}}  \, \alpha_+^a(\boldsymbol {k})  + i \cos \theta _{\ny {k}} \,  \beta_+^{a } ( \boldsymbol {k} )  \right ]  e^{- i \boldsymbol {k x} + k t }   \bigg \} \; , \cr }
 \eqno(43)
$$
\vskip -0.3 truecm
$$ 
\widehat A^a_3 (x) = \int {d^2k \over 2 \pi} \rad \bigg \{   i \beta_-^a(\ny {k} ) e^{i\boldsymbol {k  x} - kt} + i \alpha_+^{a  } ( \ny {k} ) e^{-i \boldsymbol {k  x} + kt }  \bigg \} \; , 
 \eqno(44)
$$
$$
\widehat B^a (x) = \int {d^2k \over 2 \pi} \rad \bigg \{   \alpha_-^a(\ny {k} ) e^{i\boldsymbol {k  x} - kt} - \beta_+^{a  }( \ny {k} ) e^{-i \boldsymbol {k  x}+ kt }  \bigg \} \; .  
 \eqno(45)
$$
Similarly, the ghost operators (30) and (31) become 
$$ 
\widehat {c^a} (x) = \int {d^2k \over 2 \pi} \radi2k    \bigg \{ \lambda_-^a (\ny {k})    e^{i\boldsymbol {k x} - kt } + \xi_+^a (\ny {k} )  e^{- i\boldsymbol {k x} + kt } \bigg \} \; , 
 \eqno(46)
$$
$$
\widehat  {{\overline c}^a}(x)  = \int {d^2k \over 2 \pi} \radi2k \bigg \{ -i \xi_-^a (\ny {k})    e^{i\boldsymbol {k x} - kt } + i \lambda_+^a (\ny {k} )  e^{- i\boldsymbol {k x} + kt } \bigg \}  \; .  
 \eqno(47)
$$

Let $\cal F$ be the Fock space which is canonically associated with the annihilation and creation operators  $\alpha^a_{\pm}, \beta^a_{\pm}$ and $\lambda^a_{\pm} , \xi^a_{\pm}$. The operators  $\alpha^a_{\pm}, \beta^a_{\pm}$ commute with the operators $\lambda^a_{\pm} , \xi^a_{\pm}$. The vector $| 0 \rangle \in {\cal F}$ with unitary norm, that represents the ``vacuum" state, satisfies the relations  
$$\eqalign { 
\alpha_-^a (\ny {k } )| 0 \rangle =0 \quad &, \quad \beta_-^a (\ny {k } )| 0 \rangle =0 \; , 
\cr 
\lambda_-^a (\ny {k } )| 0 \rangle =0 \quad &, \quad \xi_-^a (\ny {k } )| 0 \rangle =0 \; , \cr }  
\eqno(48)
$$
 for any $\ny {k} $ and $a$.  Expressions (42)-(47)  ---together with equations (38) and (39)---   give an explicit Fock space representation of the field operators; these operators  satisfy the free equations of motion and fulfill  the canonical commutation relations. 
 
\medskip
 
\noindent {\bf Proposition 1.} {\it In the CS model formulated in $\hbox{\matdop R}^3$, the perturbative expansion of the expectation value $\langle F \rangle $, where the functional $F = F [A^a_\mu , B^a , {\overline c}^a , c^a  ] $ admits an expansion in powers of the fields, is given by 
$$ 
\langle F \, \rangle =  { \langle 0 | \, {\rm T}  \left ( \,  F [\widehat A^a_\mu , \widehat B^a , \widehat {{\overline c}^a} , \widehat {c^a } ] \, e^{i S_I [\widehat A^a_\mu , \widehat B^a , \widehat {{\overline c}^a} , \widehat {c^a } ] } \, \right ) | 0  \rangle \over \langle 0 |\,  {\rm T} \left ( \,  e^{i S_I [\widehat A^a_\mu , \widehat B^a , \widehat {{\overline c}^a} , \widehat {c^a } ]  } \, \right ) | 0  \rangle } \; ,  
\eqno(49)
$$
where the  time-ordered product acts on the field operators $\widehat A^a_\mu , \widehat B^a , \widehat {{\overline c}^a} , \widehat {c^a } $ that are shown in equations} (42)-(47).   

\smallskip

\noindent $\underline {\rm Proof}$. Let us firstly compute the the two-point functions;  one finds 
$$\eqalign { 
\langle 0 |\,  {\rm T} \left (  \widehat c^a(x)  \, \widehat {\overline c}{^b}(y) \right ) | 0  \rangle \! &= \! i \delta^{ab}\int {d^2k \over (2 \pi )^2}  e^{i {\ny k} ( {\ny x} - {\ny y}) } {1 \over 2k} \bigg [ \theta (x^3 - y^3 ) e^{-k (x^3 - y^3) } \cr  
& {\hskip 3.8 truecm } +   \theta (y^3 - x^3 ) e^{-k (y^3 - x^3) }\bigg ] \; . \cr }
\eqno(50)
$$
\vskip - 0.7 truecm 
\noindent By means of the identity
\vskip - 0.1 truecm  
$$
\int {dk^3 \over 2 \pi } {e^{i k^3 (x^3 - y^3) } \over (k^3)^2 + k^2 } = {1\over 2k}  
\bigg [  \theta (x^3 - y^3 ) e^{-k (x^3 - y^3) } + \theta (y^3 - x^3 ) e^{-k (y^3 - x^3) }\bigg ] \; , 
\eqno(51) 
$$
\vskip - 0.2 truecm 
\noindent expression (50) can be rewritten as 
\vskip - 0.2 truecm 
$$
\langle 0 |\,  {\rm T} \left (  \widehat c^a(x)  \, \widehat {\overline c}{^b}(y) \right ) | 0  \rangle = i \delta^{ab} \int {d^3 k \over (2 \pi )^3} {e^{ik_\mu (x^\mu - y^\mu )} \over k_\mu k^\mu } \; . 
\eqno(52)
$$
\vskip - 0.3 truecm
\noindent Similarly, one gets  
$$\eqalign { 
\langle 0 |\,  {\rm T} \left (  \widehat A^a_1 (x) \widehat A^b_2 (y) \right ) | 0  \rangle \! &= \!  i \delta^{ab}\int {d^2k \over (2 \pi )^2}  e^{i {\ny k} ( {\ny x} - {\ny y}) } {1 \over 2} \bigg [ \theta (x^3 - y^3 ) e^{-k (x^3 - y^3) } \cr 
& {\hskip 3.3 truecm } -   \theta (y^3 - x^3 ) e^{-k (y^3 - x^3) }\bigg ] \cr 
&= \!  - i \delta^{ab}{\der \over \der x^3} \int {d^2k \over (2 \pi )^2}
 e^{i {\ny k} ( {\ny x} - {\ny y}) }  {1 \over 2k}  \bigg [ \theta (x^3 - y^3 ) e^{-k (x^3 - y^3) } \cr 
& {\hskip 3.8 truecm } +   \theta (y^3 - x^3 ) e^{-k (y^3 - x^3) }\bigg ] \cr 
&= \!  - i \delta^{ab}{\der \over \der x^3} \int {d^3 k \over (2 \pi )^3} {e^{ik_\mu (x^\mu - y^\mu )} \over k_\mu k^\mu }
 \; . \cr } 
\eqno(53) 
$$
More generally, it turns out that the  expectation values (49) of the time-ordered products of the couples of field operators coincide with the components (12) of the Feynman propagator, 
$$\eqalign { 
\langle 0 |\,  {\rm T} \left (  \widehat A^a_\mu (x) \widehat A^b_\nu (y) \right ) | 0  \rangle = 
\WT{A^a_\mu(x)\> A}{^b_\nu}(y)    &,    
\langle 0 |\,  {\rm T} \left (  \widehat A^a_\mu (x) \widehat B^c (y) \right ) | 0  \rangle =
 \WT{A^a_\mu(x)\> B}{^c}(y) \cr 
  \langle 0 |\,  {\rm T} \left (  \widehat B^a (x) \widehat B^c (y) \right ) | 0  \rangle =
 \WT{B^a(x)\> B}{^c}(y)   &,  
  \langle 0 |\,  {\rm T} \left (  \widehat c^a(x)  \, \widehat {\overline c}{^b}(y) \right ) | 0  \rangle =   \WT{c^a(x)\> {\overline c}}{^b}(y)  \; . \cr } 
  \eqno(54)
$$
 Consequently, since 

\item{(a)} the field operators linearly depend on the annihilation and creation operators, and
\item{(b)} the commutators (or anti-commutators) of the annihilation and creation operators are numbers (that commute with all the operators),

\noindent the expectation value  (49) of any functional of the fields reproduces precisely the perturbative expansion, which is based on the Wick contractions,  of the CS theory. 
{\hfill {\vbox{\hrule \hbox to 7 pt {\vrule height 6.2pt \hfil \vrule} \hrule}}}

\medskip 

Let us now check the basic symmetry properties of the  expectation values (49). The hamiltonian $H$ can be decomposed as $H= H_0 + g H_1$ where $H_0$ is the free hamiltonian.  By using equation (10) and the expressions of the field operators, one gets  
$$
H_0 = \int d^2 k \, (-i k) \left [ \alpha^a_+ (\ny {k } ) \alpha^a_- (\ny {k } ) + \beta^a_+ (\ny {k } ) \beta^a_- (\ny {k } ) + \xi^a_+ (\ny {k } ) \xi^a_- (\ny {k } ) +  \lambda^a_+ (\ny {k } )  \lambda^a_- (\ny {k } )  \right ]   \; ,  
\eqno(55)  
$$
\vskip - 0.3 truecm 
\noindent and then 
$$
H_0 \, | 0 \rangle = 0  \; . 
\eqno(56) 
$$
This implies invariance of the   expectation values (49) under time translations, in agreement with the statement of Proposition~1.  The annihilation and creation operators  $\alpha^a_{\pm}({\ny k})$, $ \beta^a_{\pm} ({\ny k}) $, $\lambda^a_{\pm} ({\ny k})$ and  $\xi^a_{\pm}({\ny k})$ diagonalize the time evolution which is generated by the free hamiltonian $H_0$. This means, for example, that 
$\alpha_{\pm} ({\ny k}, t)= e^{\pm kt} \, \alpha_{\pm} ({\ny k})$. A basis for $\cal F$ can be obtained by applying   products of a finite number of creation operators $\alpha^a_+({\ny k})$, $\beta^a_+({\ny p})$,   $ \lambda^a_+({\ny k})$ and $\xi^a_+ ({\ny p}) $ to $| 0 \rangle $.  Finally, let us consider the BRST charge $Q$ which can be written as $Q=Q_0 + g Q_1$ where $Q_0$ is the free component of the charge;   from equation (9) one obtains  
$$
Q_0 = \int d^2 k \, \sqrt  k \,  \left [ \xi^a_+ (\ny {k } ) \alpha_-^a (\ny {k } )   + \beta^a_+ (\ny {k } )   \lambda^a_- (\ny {k } )     \right ]    \; .  
\eqno(57)
$$
\vskip - 0.5 truecm 
\noindent Therefore
\vskip - 0.2 truecm 
$$ 
Q_0 \, | 0 \rangle = 0  \; ,  
\eqno(58) 
$$
and  the expectation values (49) are perturbatively invariant under BRST transformations, as it should be. 

The CS action terms  (6) and (7) define an euclidean quantum field theory; consistently, relations (40) and (41) show that the  field operators (42)-(45) ---that represent classical real variables--- are not hermitian. In facts, in agreement with the general properties of the field operators in the euclidean region [14],  one has 
\vskip - 0.2 truecm 
$$\eqalign { 
\left [ \widehat A^a_\mu (\boldsymbol {x}\,  , t )  \right ]^\dagger &= e^{-i t H_0} \left [ \widehat A^a_\mu (\boldsymbol {x}\,  , 0 ) \right ]^\dagger e^{i t H_0}\not=  \widehat A^{a}_\mu (\boldsymbol {x}\,  , t ) \; ,  \cr 
\left [ \widehat B^a (\boldsymbol {x}\,  , t )  \right ]^\dagger &= e^{-i t H_0} \left [ \widehat B^a (\boldsymbol {x}\,  , 0 ) \right ]^\dagger e^{i t H_0}\not=  \widehat B^{a} (\boldsymbol {x}\,  , t )  \; . \cr }
\eqno(59) 
$$
It is interesting to note that the quantum field operators  (42)-(45)  are symmetric under the following conjugation  map
\vskip - 0.3 truecm  
$$\eqalign { 
& {\hskip -0.7 truecm}  \alpha^a_- (\ny {k}) \; \mapsto \; \alpha^a_-(- \ny{k})  \quad , \quad \beta^a_- (\ny {k}) \; \mapsto \; - \beta^a_-(- \ny{k})    \cr 
& {\hskip -0.7 truecm}  \alpha^a_+ (\ny {k}) \; \mapsto \;  - \alpha^a_+(- \ny{k})  \quad , \quad  \beta^a_+ (\ny {k}) \; \mapsto \;  \beta^a_+(- \ny{k}) \; ,  \cr }     
\eqno(60) 
$$
which  ensures the reality of the eigenvalues of the field operators. The use of non-hermitian  operators in connection with real classical variables in quantum mechanics has been recently discussed in references  [15,16].

\vskip 0.6 truecm 
 
\noindent {\tmsm 4 ~ One-loop effective action in background gauge}

\medskip

\noindent The main properties of the one-loop effective action ---computed in the covariant Landau gauge shown in equations (6) and (7)---  have been discussed in reference [4]. Let us now consider the background gauge [17,18,19] which could be used to compute CS observables  in the presence of a nontrivial background field. Let us decompose the gauge fields as 
$$ 
A^a_\mu (x) = {\cal A}^a_\mu (x) + Q^a_\mu (x) \; , 
\eqno(61)
$$
where ${\cal A}^a_\mu (x) $ stands for a generic classical background configuration,  whereas $Q^a_\mu (x) $ denotes the quantum field components. The gauge-fixing part $S_{\phi \pi}$ of the action is now 
$$ 
S_{\phi \pi } = \int d^3x \left  [ (D^{ab\, \mu }({\cal A}) B^b) Q_\mu^a  + ( D^{ab \, \mu }({\cal A}) {\overline c}^b) (D^{ad}_\mu ({\cal A} + Q )  c^d )\right ] \; , 
\eqno(62)
$$
where the covariant derivative ---with respect to the background field--- is defined by 
$$ 
D^{ab}_\mu ({\cal A}) = \delta^{ab} \der_\mu  - g f^{a  c b} {\cal A}^c_\mu = 
\delta^{ab} \der_\mu  - g  {\cal A}^{ab}_\mu  \; . 
\eqno(63)
$$
Let $\Gamma [{\cal A}] $ denote the one-loop effective action in background gauge: 
$i \Gamma [{\cal A}] $ is the sum of the one-particle-irreducible vacuum-to-vacuum Feynman diagrams which are associated with the propagation of the (commuting and anticommuting)  quantum fields  in the presence of the classical background ${\cal A}^a_\mu $. 

The quadratic part $S_{{\overline c}c } $ of the action for the anticommuting variables reads  
$$
S_{{\overline c}c }  = \int d^3 x \; ( D^{ab \, \mu } ({\cal A}) {\overline c}^b ) (D^{ad}_\mu ({\cal A})  c^d  )\; , 
\eqno(64)
$$
and then the contribution $ \Gamma_g$ of the ghost fields to the effective action is given by 
$$
 \Gamma_g  = - i\,  {\rm Tr\, }^\prime \,  {\rm ln} \left [ {- D^{ab}_\mu ({\cal A})  D^{bc \, \mu }({\cal A}) \over -  \der_\mu   \der^\mu  }\right ] \; , 
 \eqno(65)
$$
where ${\rm Tr\, }^\prime $ denotes the trace operation in the orbital ($x^\mu$ or $p^\mu$) variables and the trace in the space of the adjoint representation  of the gauge group. 
 The quadratic part $S_{QB}$ of the action for the quantum variables of commuting type is 
$$
S_{QB} = \int d^3x \left  [ \, \mezzo \varepsilon^{\mu \nu \rho } Q^a_\mu  D^{ab}_\nu ({\cal A}) Q_\rho ^b +  Q_\mu^a D^{ab \, \mu} ({\cal A}) B^b \, \right ] \; .   
\eqno(66)
 $$
In terms of the complex variables 
 $$
 \psi^a =  \left [ \matrix {  \psi^a_1   \cr  \psi^a_2   \cr } \right ] = 
 \rad  \left [ \matrix { Q^a_1 + i Q^a_2   \cr  B^a - i Q_3^a \cr } \right ]   \; , 
   \eqno(67)
 $$
 $$
{\overline  \psi \, }^a =  \left [ \,  {\overline \psi\, }^a_1 , {\overline \psi \, }^a_2  \,  \right ] = 
 \rad  \left [ Q^a_1 - i Q^a_2 \, ,\,  B^a + i Q^a_3 
   \right ] \; ,  
   \eqno(68)
$$
 expression (66) can be written as 
$$ 
 S_{QB} = \int d^3x \> {\overline \psi \, }^a \left [ - i \, \gamma^\mu \, D^{ab}_\mu ({\cal A}) \right ] \psi^b 
  \; ,   
\eqno(69) 
$$
with 
$$
\gamma^1 =  \left ( \matrix { 
 0 & i   \cr  -i & 0 \cr } \right ) \; , \; \gamma^2 =  \left ( \matrix {
 0 & -1   \cr  -1 & 0 \cr } \right ) \; , \; 
 \gamma^3 =  \left ( \matrix {
 -1 & 0   \cr  0 & 1 \cr }  \right ) \; . 
 \eqno(70)
 $$
A direct computation shows that  
$$\eqalign { 
\left [ - i \, \gamma^\mu \, D_\mu ({\cal A}) \right ]^2 \! &= \!  - D_\mu ({\cal A})  D^\mu ({\cal A})\,  {\hbox{\matdop I}} + i g \varepsilon_{\mu \nu \rho }  \gamma^\mu \widetilde F^{\nu \rho} \cr 
 \! &= \! \left [ - D_\mu ({\cal A})  D^\mu ({\cal A})\right ]  \left [ {\hbox{\matdop I }} - i g \,    \Delta ({\cal A})\, \varepsilon_{\mu \nu \rho }  \gamma^\mu \widetilde F^{\nu \rho} \right ] \cr } 
 \eqno(71)
$$
where the curvature components are
$$
\widetilde  F^{ab}_{\mu \nu} = \der_\mu {\cal A}^{ab}_\nu - \der_\nu {\cal A}^{ab}_\mu - g {\cal A}^{ac}_\mu {\cal A}^{cb}_\nu + g {\cal A}^{ac}_\nu {\cal A}^{cb}_\mu \; , 
\eqno(72)
$$
and $\Delta^{ab} ({\cal A})$ is the Green function associated with the differential operator $ D^{ac}_\mu ({\cal A})  D^{cb \, \mu} ({\cal A})$, 
$$ 
D^{ac}_\mu ({\cal A})  D^{cd \, \mu} ({\cal A}) \; \Delta^{db} ({\cal A})(x,y) = \delta^{ab} \, \delta^3 (x-y) \; . 
\eqno(73) 
$$
Therefore, the contribution $\Gamma_b $ of the commuting variables to the effective action is 
$$\eqalign { 
\Gamma_b \! &= \!  i\,  {\rm Tr} \,  {\rm ln} \left [ { - i \, \gamma^\mu \, D^{ab}_\mu ({\cal A}) \over - i \, \gamma^\mu \, \der_\mu } 
\right ] = \imezzi \, {\rm Tr} \,  {\rm ln}\left [ { - i \, \gamma^\mu \, D^{ab}_\mu ({\cal A}) \over - i \, \gamma^\mu \, \der_\mu } 
\right ]^2 \cr 
&= \! i\,  {\rm Tr\, }^\prime \,  {\rm ln} \left [ {- D^{ab}_\mu ({\cal A})  D^{bc \, \mu }({\cal A}) \over -  \der_\mu  \der^\mu }\right ]  + \imezzi \,  {\rm Tr} \,  {\rm ln}   \left [ {\hbox {\matdop I} } - i g \,    \Delta ({\cal A})\, \varepsilon_{\mu \nu \rho }  \gamma^\mu \widetilde F^{\nu \rho} \right ] \; , 
\cr } \eqno(74) 
$$
where the trace operation denoted by $ {\rm Tr} $ now includes (in addition to the sum over the orbital and gauge group variables) the trace  in the two-dimensional space where the components of the fields $\psi^a$ and ${\overline \psi \,}^{a}$ live. 

If  one adopts the same regularization prescription for the divergent diagrams which are associated with the ghost propagating fields and with the vector propagating fields, the contribution (65) of the ghosts cancels with part of the gauge vectors contribution (74), and the  renormalized effective action $\Gamma = \Gamma_g + \Gamma_b $ in background gauge  takes the form
$$
\Gamma [{\cal A}] =  \imezzi \,  {\rm Tr} \,  {\rm ln}   \left [ {\hbox{\matdop I} } - i g \,    \Delta ({\cal A})\, \varepsilon_{\mu \nu \rho }  \gamma^\mu \widetilde F^{\nu \rho} \right ] \; . 
\eqno(75) 
$$
Expression (75) is perturbatively well defined; each nontrivial term of its Taylor expansion in powers of ${\cal A}^a_\mu (x)$ is a finite functional of the background field (it does not contain ultraviolet divergences). 
$\Gamma [{\cal A}]  $ is  a manifestly gauge-invariant functional of the classical configuration ${\cal A}^a_\mu (x)$, as it should be in the background gauge formalism.  Finally, expression (75) shows that $\Gamma [{\cal A}]  $ vanishes when the curvature components ---that are associated with the configuration ${\cal A}^a_\mu (x)$--- are vanishing.

\vskip 1.1 truecm 

 \noindent {\bf Acknowledgments.}  I wish to thank G. Cicogna, L.E. Picasso  and R. Stora  for useful discussions. 

\vskip 1.5 truecm 

\noindent {\tmsm References} 

\vskip 0.7 truecm 

\item {[1]} A.S.~Schwarz, Lett. Math. Phys. 2 (1978) 247.  

\medskip 

\item {[2]}  A.S.~Schwarz, Commun. Math. Phys. 67 (1979) 1.  

\medskip

\item {[3]} E.~Witten, Commun. Math. Phys. 121 (1989) 351. 

\medskip

\item {[4]} E.~Guadagnini, M.~Martellini and M.~Mintchev, Phys. Lett. B  227 (1989) 111. 

\medskip

\item {[5]} E.~Guadagnini, M.~Martellini and M.~Mintchev, Nucl. Phys. B330 (1990) 575. 

\medskip

\item {[6]} A.C. Hirshfeld and U. Sassenberg, Journal of Knot Theory and its Ramifications, 5 (1996) 805. 

\medskip

\item {[7]} F.J.~Dyson, Phys. Rev. 75 (1949) 486.

\medskip

\item {[8]} E.~Guadagnini, {\it Functional integration and abelian link invariants}, Studies in Advanced Mathematics ``Chern-Simons Gauge theory: 20 years after", edited by J.E.~Anderson, H.~Boden, A.~Hahn and B.~Himpel, e-print arXiv:1001.4645v1. 

\medskip

\item {[9]} F. Thuillier, J. Math. Phys. 50, 122301 (2009); e-print arXiv:0901.2485.

\medskip

\item {[10]} N.N.~Bogoliubov et D.V.~Chirkov, {\sl Introduction a la th\'eorie quantique des champs\/}, Dunod  Editeur (Paris, 1960). 

\medskip

\item {[11]}  F. Delduc, O. Piguet, C. Lucchesi and S.P. Sorella, Nucl. Phys. 
B 346 (1990) 313. 

\medskip

\item {[12]}  E.~Guadagnini, {\sl The Link Invariants of the Chern-Simons Field Theory}, 
De Gruyter Expositions in Mathematics, Vol.~10, edited by O.H.~Kegel, V.P.~Maslov, W.D.~Neumann and R.O.~Wells Jr. (Walter De Gruyter, Berlin, 1993).

\medskip

\item {[13]} D.~Deturck, H.~Gluck, R.~Komendarczyk, P. Melvin, C.~Shonkwiler and D. S. Vela-Vick, {\it Pontryagin invariants and integral formulas for Milnor's triple linking number}, e-print arXiv:1101.3374v1, 18 Jan 2011.  

\medskip

\item {[14]} K.~Symanzik, J. Math. Phys. 7 (1966) 510.  

\medskip

\item {[15]} A. Mostafazadeh, {\it Conceptual Aspects of ${\cal PT}$-Symmetry and Pseudo-Hermiticity: A status report}, arXiv:1008.4680v1,  27 Aug 2010. 

\medskip

\item {[16]} K.~Jones-Smith and H.~Mathur, {\it A New Class of non-Hermitian Quantum Hamiltonians with ${\cal PT}$ symmetry}, arXiv:0908.4255v4,  4 Oct 2010. 

\medskip

\item {[17]} J.~Honerkamp, Nucl. Phys. B 36 (1971) 130.  

\medskip

\item {[18]} G.~'t Hooft, Nucl. Phys. B 62  (1973) 444.  

\medskip

\item {[19]} L.F.~Abbott, Nucl. Phys. B  185 (1981) 189.

 \vfill\eject 
\end 
\bye